\documentclass[aps,prl,reprint,showpacs,superscriptaddress]{revtex4-2}
\usepackage{graphicx}
\usepackage{amsmath}
\usepackage{amssymb}
\usepackage{amsfonts}
\usepackage{dcolumn}
\usepackage{xcolor}
\usepackage{subfigure}
\usepackage{bm}
\usepackage{amsthm}
\usepackage[T1]{fontenc}
\usepackage[mathscr]{euscript}
\DeclareMathAlphabet{\mathcalligra}{T1}{calligra}{m}{n}
\usepackage{dcolumn}
\usepackage{capt-of}
\usepackage{pifont}
\usepackage{titlesec}
\titleformat{\section}{\Large\bfseries\centering}{\thesection}{1em}{}
\titleformat{\subsection}{\normalsize\bfseries\centering}{\thesubsection}{1em}{}
\titleformat{\subsubsection}{\normalsize\bfseries\centering}{\thesubsubsection}{1em}{}
\usepackage{hyperref}
\hypersetup{
colorlinks=true,final=true,
        linkcolor=blue,
        citecolor=blue,
        filecolor=blue,
        urlcolor=magenta,
}

\usepackage{bigints}   
\usepackage{epstopdf}   
\usepackage{soul} 

\begin{document}
\title{Chiral soft mode transition driven by strain in ferroelectric bubble domains}
\author{Urmimala Dey}
\email{urmimala.dey@list.lu}
\affiliation{Smart Materials Unit, Luxembourg Institute of Science and Technology (LIST), Avenue des Hauts-Fourneaux 5, L4362, Esch-sur-Alzette, Luxembourg}
\author{Natalya S. Fedorova}
\affiliation{Smart Materials Unit, Luxembourg Institute of Science and Technology (LIST), Avenue des Hauts-Fourneaux 5, L4362, Esch-sur-Alzette, Luxembourg}
\author{Jorge \'{I}\~{n}iguez-Gonz\'{a}lez}
\affiliation{Smart Materials Unit, Luxembourg Institute of Science and Technology (LIST), Avenue des Hauts-Fourneaux 5, L4362, Esch-sur-Alzette, Luxembourg}
\affiliation{Department of Physics and Materials Science, University of Luxembourg, 41 Rue du Brill, L4422, Belvaux, Luxembourg}
\author{Hugo Aramberri}
\email{hugo.aramberri@list.lu}
\affiliation{Smart Materials Unit, Luxembourg Institute of Science and Technology (LIST), Avenue des Hauts-Fourneaux 5, L4362, Esch-sur-Alzette, Luxembourg}
\date{\today}

\begin{abstract}
Chirality in solids is attracting growing attention as a potential ferroic order, yet virtually no paradigmatic example of a soft-mode achiral-to-chiral phase transition has been firmly established to date. Here we identify ferroelectric bubble domains as a model system that undergoes a strain-driven achiral-to-chiral transition exhibiting the hallmarks of spontaneous symmetry breaking. Using second-principles atomistic simulations, we uncover chiral phonon modes in ferroelectric/dielectric superlattices that soften under epitaxial strain following textbook soft-mode behaviour. The transition is accompanied by a change in topological character, highlighting an interplay between chirality and topology in these systems. This work provides a concrete step towards establishing chirality as a genuine ferroic order in solids.
\end{abstract}

\maketitle

A renewed interest in chirality has sprouted in materials science. On the one hand, control over chirality has the potential to unlock a new ferroic order, which could come together with many promising technological applications, including the possibility to enhance molecular enantioselectivity, which is key to drug synthesis in pharmaceutical development or agrochemistry~\cite{knowles02,mondal22} among other industries. On the other hand, it is still not clear how to measure chirality for most chiral solids~\cite{weinberg00,kishine22,gomezortiz24}, or if handedness could be unequivocally defined for every chiral object~\cite{ruch72,king03,fecher22}. Thus, it remains unclear what the associated conjugate field and susceptibility would even be, let alone whether there is a foolproof recipe to define a chiral order parameter~\cite{vavilin22,kishine22,oiwa22,bousquet25}, which poses stimulating challenges.

Chiral phonons are lattice vibrations that endow an otherwise achiral system with chirality~\cite{zhang15,zhu18,romao24}. Because of the dynamic nature of phonons, such chirality is often only a transient or instantaneous one that vanishes upon time-averaging. Some strategies (based on non-linear phonon dynamics~\cite{juraschek17,romao24,zeng25a}, or on the application of very large magnetic fields~\cite{juraschek19}) have been used to turn an achiral system into a chiral one under a certain stimulus. 

A soft-mode-driven chiral phase transition would correspond to a system consolidating the instantaneous chirality of the phonon and with a particular handedness, as is characteristic of spontaneous symmetry-breaking transitions. Interestingly, only a handful of achiral-to-chiral structural phase transitions have been reported~\cite{hirotsu75,ohsato90,schmidt04,djurivs12,kimura16,zeiger25}, and even fewer works have shown an achiral phase with chiral instabilities~\cite{fava24,fava25,gomezortiz25,zheng25}. Only very recently two examples of achiral-to-chiral transition driven by a phonon soft-mode have been predicted~\cite{fava25,zheng25}. The scarcity of well-described examples of such transitions significantly hinders the recognition of chirality as a legitimate ferroic order.

\begin{figure}[ht]
\includegraphics[scale=0.84]{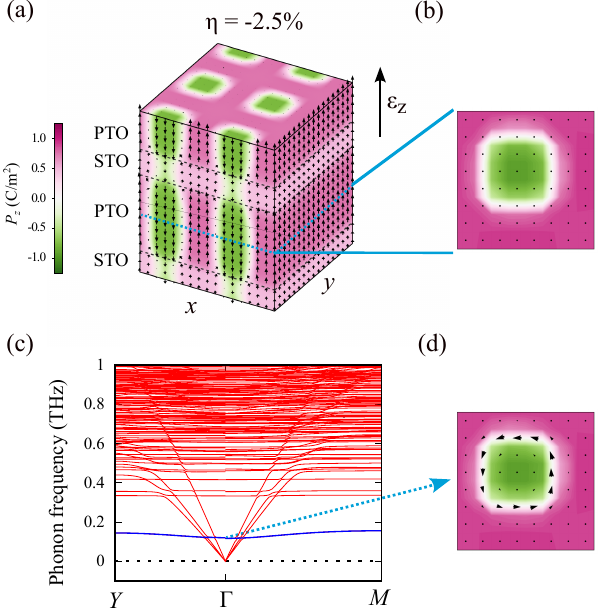}
    \caption{(a) 3D view of the electric bubble lattice in a ${\rm{(PTO)}}_9$/$\rm{{(STO)}}_3$ superlattice under in-plane compressive strain $\eta$. Each arrow indicates the electric dipole associated with a single 5-atom perovskite unit cell. (b) Dipole field in the middle PTO layer. (c) Low-frequency phonon spectrum computed for the bubble lattice shown in (a). The chiral phonon branch is highlighted in dark blue. (d) The eigenmode of the lowest frequency optical phonon displays a Bloch component all along the bubble's domain wall.}
\label{fig1}
\end{figure}

In this work, we predict that ferroelectric bubble domains in perovskite ferroelectric/dielectric heterostructures undergo an achiral-to-chiral phase transition driven by a soft mode induced by epitaxial strain. Using first-principles-based atomistic models, we show that this transition provides a platform to study spontaneous symmetry-breaking chiral phase transitions in solids. Furthermore, we find that the handedness can exhibit spatial modulation not only among bubble domains, but also within individual bubbles, allowing for the study of different arrangements of chiral order.

Ferroelectric materials under open-circuit electric boundary conditions become frustrated. The absence of a screening charge at the boundaries often results in the ferroelectric breaking into domains of opposite polarization separated by a domain wall~\cite{aguado12,wojdel14,yadav16}.
Ferroelectric/dielectric heterostructures, and in particular PbTiO$_3$/SrTiO$_3$ (PTO/STO) superlattices, have become a rich playground for exploring topological defects, nanoscale domain architectures, and emergent functionalities driven by domain wall phenomena~\cite{junquera23}.
In this scenario, an imbalance between the up and down domains (be it a systemic asymmetry or an applied field) can result in the formation of minority cylindrical bubble domains in the ferroelectric layer~\cite{lichtensteiger14,das19,han22,aramberri24} [see Fig.~\ref{fig1}(a)]. Such domains have been observed to be $5-7$~nm in cross section diameter~\cite{das19,han22}, and according to predictions by some of us they could be as small as a handful of unit cells~\cite{aramberri24}. The bubble domains are expected to span the whole thickness of a ferroelectric thin film.

In Fig.~\ref{fig1}(a), we show the dipole field for a bubble lattice in a PTO/STO superlattice. Under compressive epitaxial strain, the walls enclosing the bubble display an Ising character [see Fig.~\ref{fig1}(b)], except at the layers adjacent to the interface with STO, which present Néel character. The resulting structure is achiral (space group $P4mm$ for the bubble arrangement in the figure).

In Fig.~\ref{fig1}(c) we show the phonon spectrum computed for the bubble lattice in Fig.~\ref{fig1}(a). The lowest frequency optical mode at the zone center corresponds to the appearance of Bloch components at the bubble wall [see Fig.~\ref{fig1}(d)]. These wall dipoles are essentially tangential to the wall and form a vortex. Note that changing the sign of the mode amplitude reverses the vortex handedness. The resulting structure loses all mirror symmetries and becomes clearly chiral (space group $P4$), with the handedness determined by the sign of the mode amplitude. Strictly speaking, the phonon eigenvector itself is inversion-symmetric and therefore not chiral. However, since inversion symmetry is already broken in the bubble lattice, the mode further breaks the remaining improper symmetries, thereby rendering the structure chiral. Therefore, the lowest frequency phonon at the $\Gamma$-point is a (structural) chiral phonon.

In ferroelectric perovskites like PTO, it is known that at ferroelectric domain walls a Bloch dipole component may appear~\cite{stepkova12,wojdel14}. Since it condenses as a secondary distortion, it vanishes with increasing temperature before the material becomes paraelectric and the domains (and walls) themselves disappear~\cite{wojdel14,aramberri24}. Moreover, owing to the strong strain–polarization coupling of ferroelectric perovskites, the Bloch component is suppressed under compressive strain~\cite{stepkova12,aramberri24,zatterin24}.
In contrast, when the system is driven towards a more tensile epitaxial strain regime, the Bloch component is stabilized. We also observe this in the ferroelectric bubble domains:
If we take the achiral bubble shown in Fig.~\ref{fig1}(a) to 0\% strain, the vortical Bloch dipoles condense spontaneously upon structural optimization, clockwise and anticlockwise vortices being degenerate energy minima, as seen in Fig.~\ref{fig1}(d). 

\begin{figure}[ht]
\includegraphics[scale=0.47]{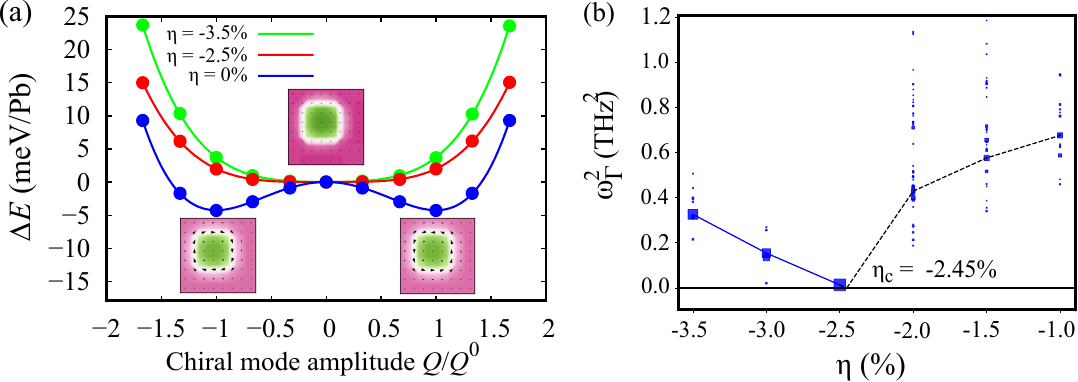}
    \caption{(a) Energy landscape of the ferroelectric bubble lattice as a function of the chiral mode amplitude $Q$ under different epitaxial strains, showing the transition from a single-well (achiral) to a double-well (chiral) potential in the tensile regime. The chiral mode amplitude is scaled to its relaxed magnitude $Q_0$ at 0\% strain. (b) Squared frequency of the chiral phonon at the $\Gamma$ point as a function of epitaxial strain, illustrating the linear softening towards the critical strain ($\eta_c$) characteristic of a soft-mode-driven phase transition.}
\label{fig2}
\end{figure}

In Fig.~\ref{fig2}(a), we show the energy of the bubble lattice as a function of the chiral phonon amplitude for varying in-plane epitaxial strain. For compressive strains ($\eta = -3.5\%$ and $-2.5\%$), the energy profile shows a single minimum corresponding to the achiral structure, the profile being flatter for the less compressive strain. By contrast, in the tensile regime ($\eta = 0\%$), the chiral mode becomes an instability and the energy profile takes the form of a double well, characteristic of spontaneous symmetry-breaking ferroic transitions. Thus, our results demonstrate that strain drives the achiral-to-chiral phase transition of the polar bubbles.

Next, we monitor the frequency of the chiral mode as a function of epitaxial strain, as shown in Fig.~\ref{fig2}(b). We find an essentially linear evolution of the squared chiral phonon frequency down to the critical strain ($\eta_{c} \approx -2.45\%$). Although the chiral mode mixes with other symmetry-compatible modes, the mode with the largest projection onto the chiral phonon eigenvector still follows a roughly linear trend with strain near the critical point. We thus find the textbook behaviour of a second-order phase transition driven by a soft-mode~\cite{lines01}.

Note that the strain-driven achiral-to-chiral transition of the bubble domains comes hand in hand with a topological phase transition. In the achiral state, the polarization texture across the bubble wall is purely Ising-like and carries a trivial skyrmion number~\cite{das19,junquera23}, while in the chiral phase the emergence of a Bloch component endows the wall with a non-zero winding and thus a finite skyrmion number $N_{\rm{sk}} = +1$. The transition therefore corresponds to a change in the topological charge of the bubble~\cite{pereira19}, driven by the softening of the chiral phonon mode. This highlights the connection between topology and chirality in ferroelectric bubbles, where the skyrmion number and the chiral ferroic order evolve together across the soft-mode transition.

Interestingly, we find that the strain-driven chiral phase transition of electric bubble lattices is qualitatively independent of the bubble size. For example, we repeated the above analysis for a bubble lattice consisting of bubbles with a radius of a single perovskite unit cell (see Supplementary Fig.~S2). We again observe a strain-driven transition from a single-well to a double-well energy profile, together with a roughly linear behaviour of the squared frequency near the critical strain ($\eta_{c} \approx -1\%$), showing that the soft-mode-driven second-order phase transition of electric bubbles is a general behaviour (see Supplementary Fig.~S8). For computational convenience and to make the analysis simpler, we work with these smaller electric bubbles in what follows.

Several factors can influence the chiral phonon frequencies and ultimately the critical strain. For instance, we find that thicker (thinner) PTO layers shift the critical strain to less (more) compressive values (see Supplementary Fig.~S9), consistent with the fact that thicker ferroelectric layers exhibit larger out-of-plane polarization, which reduces the critical strain required to suppress the Bloch components. 

We also compute the effect of the applied electric field $\mathcal{E}_z$ on the chiral phonon frequency at a fixed strain of $-2\%$. We find that the vertical field acts as an effective additional compressive strain, causing the chiral mode to harden by about 0.1~THz when $\mathcal{E}_z$ is increased by approximately 750~kV~cm$^{-1}$ (see Supplementary Fig.~S12).  The frequency change is linear with the applied field, unlike with strain, which induces a linear change in the squared frequency (see Supplementary Note~S2 for more details on this). Hence, the vertical electric field provides an additional knob to fine-tune the chiral phonon mode.

\begin{figure}[ht]
\includegraphics[scale=1.05]{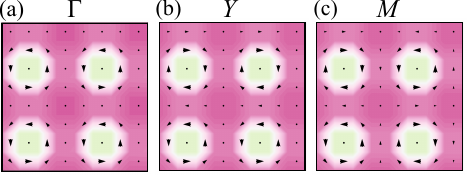}
    \caption{Bubble lattice structures obtained by condensing the chiral soft mode at different points of the Brillouin zone: (a) $\Gamma$-point (uniform chirality, ferrochiral order), (b) $Y$-point, and (c) $M$-point modulations. The $Y$ and $M$ modes yield  arrays of bubbles with alternating handedness (antiferrochiral order).}
\label{fig3}
\end{figure}

Having established the existence of a soft-mode-driven achiral-to-chiral transition, we next examine how this chiral order may spatially organize across the bubble lattice. The chiral phonon branch (highlighted in blue in Fig.~\ref{fig2}(b) and also in Supplementary Fig.~S3 for a denser bubble lattice) is rather flat, already suggesting a low energy cost for spatially modulated versions of chiral bubble structures. To test this, we condense the modulated chiral soft modes at the zone-boundary $Y$ and $M$ points of the Brillouin zone (Fig.~\ref{fig3}), and find that, upon structural optimization, their energies are slightly lower than that of the non-modulated chiral structure ($\Gamma$-point mode). At $0\%$ strain, the $Y$-point ($M$-point) modulation is $-0.10$~meV/f.u. ( $-0.34$~meV/f.u.) lower in energy (see Supplementary Table~S1 for the evolution of these energy differences with strain). Note that the spatial modulation of chiral bubbles along $y$ (phonon at the $Y$ point) is equivalent by symmetry to that along $x$. This indicates that arrays of chiral bubbles with alternating handedness (\textit{antiferrochiral} order) are slightly preferred over a uniform (\textit{ferrochiral}) arrangement in which all bubbles share the same handedness.

The observed preference for alternating handedness can be rationalized by considering the relative orientation of the parietal dipoles in neighbouring bubbles. In a pair of chiral bubbles with opposite handedness, the wall dipoles closest to each other point in the same in-plane direction. The less costly it is for the matrix between bubbles to develop an in-plane dipole (i.e., under more tensile strain), the more the modulation of handedness is favoured. This coupling between neighbouring bubble walls underlies the weak energetic bias towards antiferrochiral order.

Let us stress here that the bubble-bubble interactions are likely overestimated here compared to more realistic scenarios. Our simulation box (which we choose for computational feasibility) is rather small, leading to very short inter-bubble distances which amplify the relatively small bubble-bubble interactions~\cite{aramberri24}. 

Overall, however, these arrangements are nearly degenerate in energy. Our calculations show no preference for either handedness within a single bubble (left- and right-handed bubbles are degenerate) and only a minor preference for in-plane modulation of handedness, indicating that, under no handedness-selecting stimulus, chiral bubbles of both handedness will likely form spontaneously, most likely in a disordered racemic mixture.

\begin{figure}[ht!]
\includegraphics[scale=0.48]{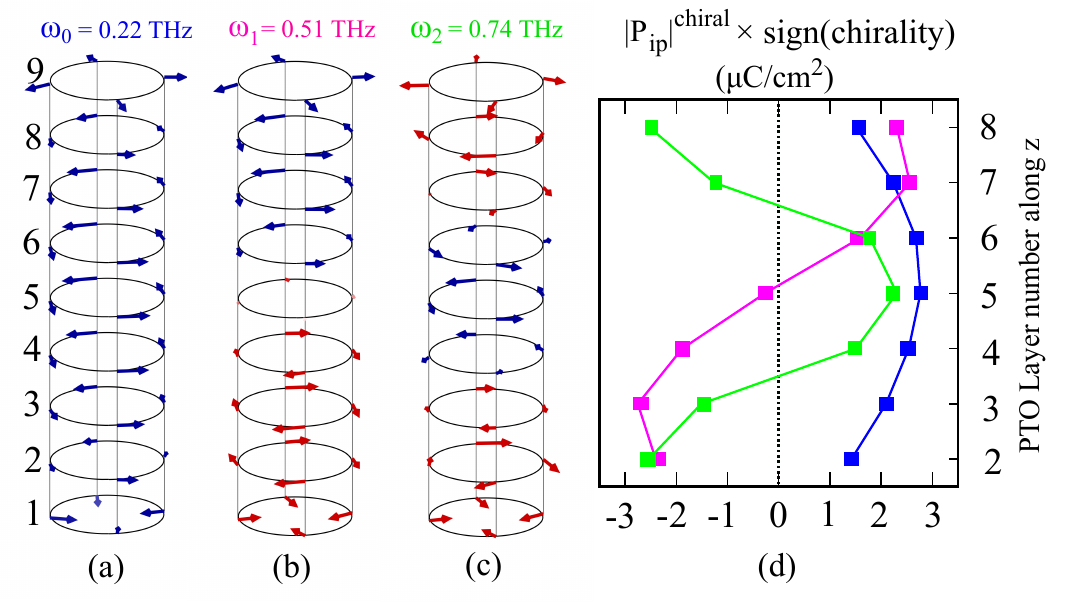}
    \caption{Vertically modulated chiral bubble structures. (a) Chiral bubble with uniform handedness. (b–c) Vertically modulated chirality with one and two Ising rings across the PTO layers, respectively, obtained from higher-frequency eigenmodes. Dipole field only at the domain walls is shown for clarity. (d) Layer-resolved in-plane polarization magnitude, showing nodal patterns analogous to harmonic modes.}
\label{fig4}
\end{figure}

Beyond these in-plane modulations, our calculations reveal an even more exotic form of chiral ordering: a vertical modulation of the handedness within a single bubble. While the chiral phonon discussed so far exhibits a uniform handedness across the bubble height [see Fig.~\ref{fig4}(a)], we also find higher-frequency modes in which the bubble cross-section still shows a chiral wall, but the handedness changes sign across different PTO layers [see Figs.~\ref{fig4}(b),(c) and Supplementary Fig.~S3]. These vertical modulations of the chiral phonon mode closely resemble the harmonic modes of a string fixed at both ends [see Fig.~\ref{fig4}(d)]. It is worth noting that at the zone-boundary, these modes involve modulations within individual bubbles as well as across neighbouring bubbles. In the PTO layers where the chirality changes sign, the Bloch component vanishes and the walls are (locally) Ising-like. Such layers bear resemblance with the extended defects known as Bloch lines, which constitute a domain wall within the domain wall. However, unlike the Bloch lines originally described for ferromagnetic systems~\cite{slonczewski74}, here the dipoles do not rotate but instead vanish at the defect. Hence, we suggest that the term \textit{Ising rings} (by analogy with the Ising lines in ferroelectrics~\cite{stepkova15}) is more appropriate in this context.

To explore their stability, we track the evolution of these modes at $\Gamma$ with strain, finding that they closely follow the behaviour of the chiral soft mode (see Supplementary Fig.~S10). Yet, we do not find a regime in which they fall below the uniformly chiral bubble in frequency. We have also tried to stabilize such structures without success. Our structural optimizations starting from the vertically modulated chiral eigenmodes typically relax into uniformly chiral bubbles. Nevertheless, the low frequencies of the vertically modulated configurations (within 0.5~THz, see Supplementary Fig.~S3) suggest that they lie close in energy to the ground state. Hence, if the ground-state order were precluded (for example, by a local compressive strain in the central PTO layer induced by point or other defects), these structures could condense.

We have also explored whether other low-energy lattice modes might compete with the chiral instability. By analyzing other phonon excitations of the achiral bubbles, we find that the optical mode closest in frequency to the chiral phonon at the $\Gamma$ point often corresponds to a type-II bubble~\cite{pereira19}, i.e., an excitation with a net in-plane dipole around the domain wall, as shown in Supplementary Fig.~S13. Given its proximity in frequency, this configuration may compete energetically with the chiral bubble phase (see Supplementary Note~S3 for more details). We further find that this mode also presents harmonics with vertical modulations of the in-plane dipole texture (see Supplementary Fig.~S14), analogous to the modulations observed in chiral bubbles. 

Temperature provides another natural control parameter for chirality in these systems. The Bloch component in the domain wall has been predicted to have a lower Curie point than that of the bulk for pure PbTiO$_3$~\cite{wojdel14,zatterin24}, and also for ferroelectric bubble domains in PTO/STO~\cite{gomezortiz22,aramberri24}. Therefore, while here we present the soft-mode behaviour with epitaxial strain, we also expect a soft-mode chiral-to-achiral transition with temperature, which could be experimentally more accessible. Explicit temperature-dependent simulations, while beyond the scope of this work, could therefore test this expectation.

From a practical perspective, measuring chirality in ferroelectric bubbles poses an experimental and even a theoretical challenge and remains an open problem~\cite{fecher22,bousquet25}. Yet, chiral objects are known to display optical activity~\cite{shafer18,junquera23,luk24} and second-harmonic-generation-based circular dichroism~\cite{das19,behera22,junquera23,luk24}. Although such responses are not unique to chiral systems, and in non-enantiomorphic space groups like the one at hand the signal can be orientation-dependent~\cite{banik16}, we believe both techniques could be used to experimentally verify our predictions. The chiral phonon is predicted to lie at a sub-THz frequency in the strain range considered here and may harden somewhat in the achiral regime at finite temperatures—still well within the reach of infrared and THz time-domain spectroscopy.

More importantly, recent experiments show that chirality can be induced in solid systems using phonon rectification~\cite{zeng25a} or even switched with conjugate fields involving circularly polarized THz pulses~\cite{zeng25b}. Theoretical works have further predicted optical control and reversal of skyrmion chirality in ferroelectric thin films~\cite{luk24,gao24}. Compared to polar vortices, bubble domains require the reversal of far fewer dipoles to switch chirality, which suggests the process would be significantly less costly~\cite{behera22,junquera23}. We thus believe bubble domains could serve as an ideal platform for such explorations.

Our findings also connect with a broader body of recent work on chiral structures in PTO/STO superlattices~\cite{shafer18,behera22,gomezortiz22,susarla23}, including a theoretical study based on a (non-atomistic) phase-field perturbation model in which strain is shown to drive the achiral-to-chiral transition~\cite{zheng25}. In all these studies, the chirality is associated with the appearance of the Bloch components at the walls of ferroelectric stripe domains, which are virtually infinitely long. This extended nature of the chiral distortion can be advantageous for obtaining stronger signals, although in real samples mobile Bloch (or Ising) lines may appear at the walls, which could make their chiral character difficult to stabilize. The bubbles considered in this study, on the contrary, are finite-size domains that can host localized chirality. Their reduced dimensionality may make them harder to measure, but might render their switching less costly, and potentially enable both information storage and transport at much higher densities.

In summary, we have shown that ferroelectric bubble domains present a chiral-to-achiral phase transition as a function of epitaxial strain driven by a chiral soft-mode. This constitutes one of the first clear examples of such a transition, which could serve as a playground to utilize chirality as a ferroic order. Importantly, this transition also has a topological character, highlighting the connection between topology and chiral ferroic order.  Moreover, we have found energy-competitive modulations of the chirality among and within the bubbles, which could further serve to enhance functionality in future devices based on handedness as an order parameter.

Work funded by the Luxembourg National Research Fund (FNR) through grant C23/MS/17909853/BUBBLACED.\\

\textit{Appendix: Simulation Details}.— We performed atomistic second-principles simulations using the SCALE-UP package~\cite{wojdel13,Garcia16,Escorihuela17} employing models of PbTiO$_3$/SrTiO$_3$ superlattices described in detail in previous studies. The superlattice models were derived from previously validated bulk models of PbTiO$_3$ and SrTiO$_3$, which were fitted to first-principles data and modified for superlattice configurations~\cite{wojdel13,wojdel14,Bellido17,zubko16}. These models were shown to accurately capture the lattice dynamical properties of both the individual materials and their superlattices~\cite{zubko16,aramberri22}. In our work, we used $8 \times 8 \times 1$ ($4 \times 4 \times 1$) simulation boxes with periodic boundary conditions to simulate the bubble lattice with bigger (smaller) bubbles. Models with varying thicknesses of the PbTiO$_3$ layers (of 6, 9 and 12 perovskite unit cells) were considered to assess the generality of our results (see Supplementary Fig.~S9). An SrTiO$_3$ substrate is taken as the zero strain ($\eta$ = 0\%) reference. 

Structural optimizations were performed via Monte Carlo annealings with an annealing rate of 0.9975 for 50,000 sweeps. The initial temperature was set to 10~K. 

In order to stabilize the bubble domains, we applied an electric field of $\mathcal{E}_z$=1.5~MV~cm$^{-1}$ for the smaller bubbles. We found that the bigger bubbles considered are stable under no applied field.

Phonopy was used to plot the dispersion curves and generate the modulated structures~\cite{togo15}. A $2 \times 2 \times 1$ supercell was used for phonon calculations. To investigate the effects of in-plane compressive strain and vertical electric field on the phonon modes, structural optimization (via Monte Carlo annealing) was performed for each strain and field condition prior to the phonon calculations. To account for the long-range electrostatic interactions in the phonon spectra near the zone center, a non-analytical correction was applied to the dynamical matrices~\cite{gonze97}, as in previous works~\cite{aramberri22}.


%
\clearpage
\newpage
\onecolumngrid
\section*{Supplementary Material for Chiral soft mode transition driven by strain in ferroelectric bubble domains}
\setcounter{page}{1}
\setcounter{figure}{0}
\setcounter{table}{0}
\setcounter{section}{0}
\renewcommand{\thepage}{S\arabic{page}}
\renewcommand{\thesection}{S\arabic{section}}
\renewcommand{\thetable}{S\arabic{table}}
\renewcommand{\thefigure}{S\arabic{figure}}
\newcounter{SIfig}
\renewcommand{\theSIfig}{S\arabic{SIfig}}



\subsection*{S1. Effect of Compressive Epitaxial Strain}
\subsubsection*{S1.1. Bigger Bubble}
\begin{figure}[ht!]
\includegraphics[scale=1.75]{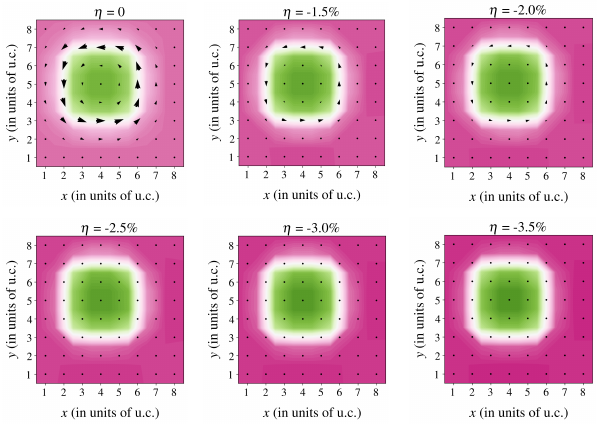}
    \caption{Evolution of the dipole field in the bulk-like PbTiO$_3$ (PTO) layer for bubble domains with a $3 \times 3$ perovskite u.c. square section, as a function of epitaxial strain $\eta$ in the 9PTO/3STO superlattice without an external electric field. The colourbar representing the magnitude of out-of-plane polarization follows the same scale as in Fig.~1(a) of the main text.}
\label{dipole_tex_strain_9_3}
\end{figure}

\newpage
\subsubsection*{S1.2. Smaller Bubble}
\begin{figure}[ht!]
\includegraphics[scale=1.4]{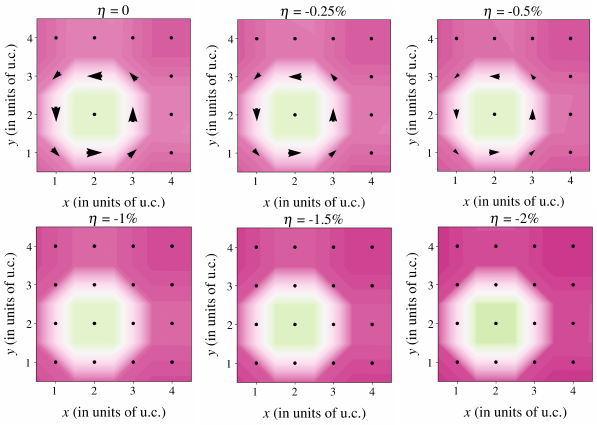}
    \caption{Evolution of the dipole field in the bulk-like PTO layer for bubble domains with a $1 \times 1$ perovskite u.c. square section, as a function of epitaxial strain $\eta$ in the 9PTO/3STO superlattice at a vertical field of $\mathcal{E}_z = 1.5$ MV/cm.}
\label{dipole_tex_strain_9_3_s}
\end{figure}

\begin{figure}[ht!]
\includegraphics[scale=0.55]{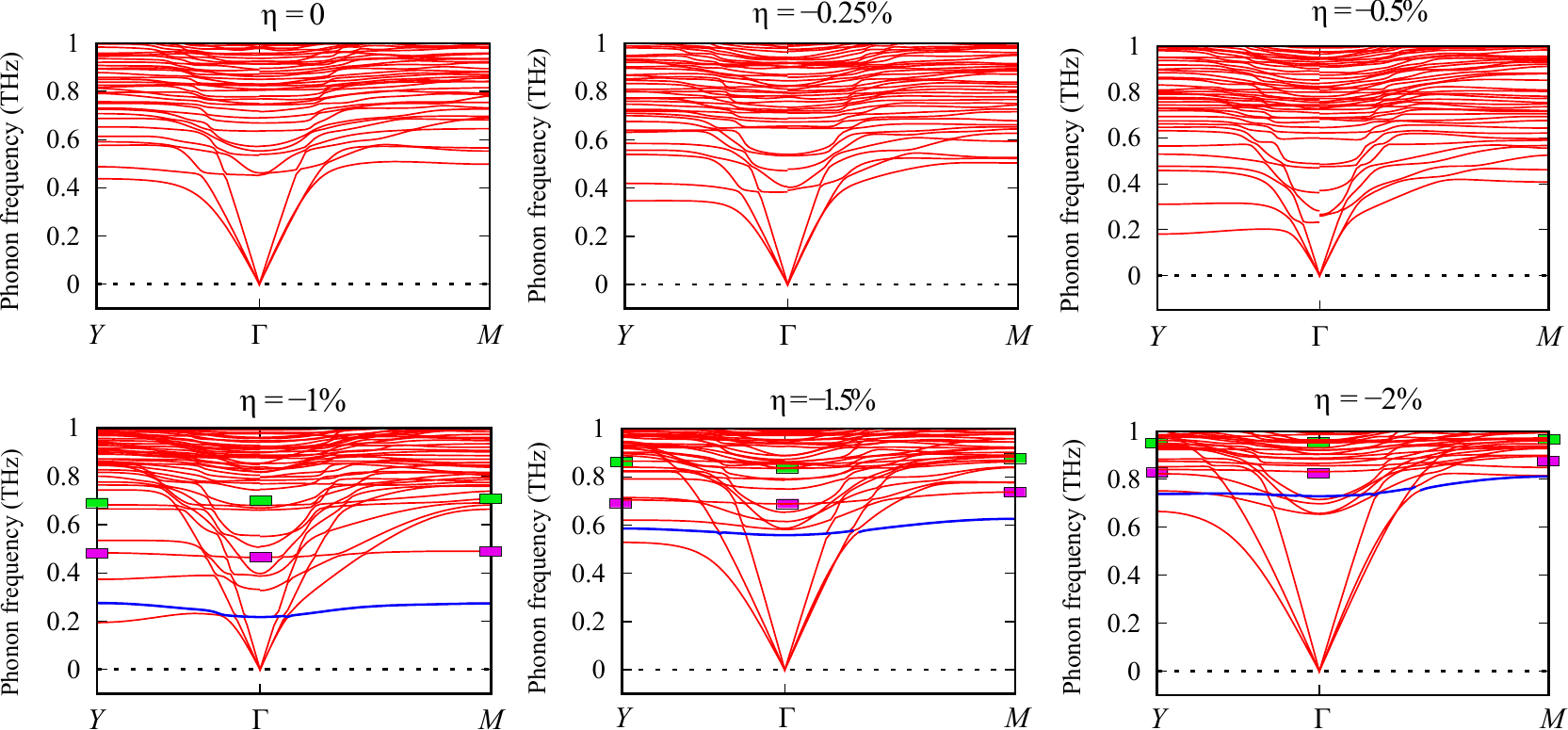}
    \caption{Evolution of the phonon band structure with epitaxial strain $\eta$ in the 9PTO/3STO superlattice at a vertical field of $\mathcal{E}_z = 1.5$ MV/cm. The chiral phonon branch, highlighted in blue, corresponds to uniform Bloch handedness throughout the thickness of the PTO layers. For $\eta < \eta_c ~(-0.92\%$), the chiral mode mixes with other symmetry-compatible modes, making it difficult to identify the chiral phonon branch unambiguously. The magenta and green rectangles indicate phonon frequencies corresponding to vertically modulated chiral structures at different high-symmetry points, featuring one and two Ising rings within the PTO block, respectively. For clarity, phonon bands are shown only up to 1 THz.}
\label{phonon_strain_9_3}
\end{figure}

\begin{figure}[ht!]
\includegraphics[scale=1.75]{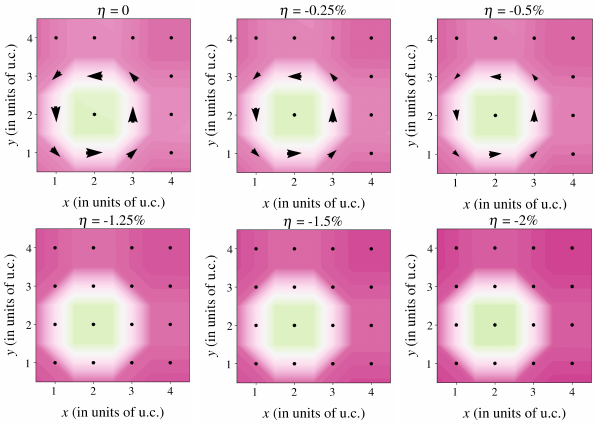}
    \caption{Evolution of the dipole field in the bulk-like PTO layer with epitaxial strain $\eta$ in the 6PTO/3STO superlattice at a vertical field of $\mathcal{E}_z = 1.5$ MV/cm.}
\label{dipole_tex_strain_6_3}
\end{figure}

\begin{figure}[ht!]
\includegraphics[scale=0.7]{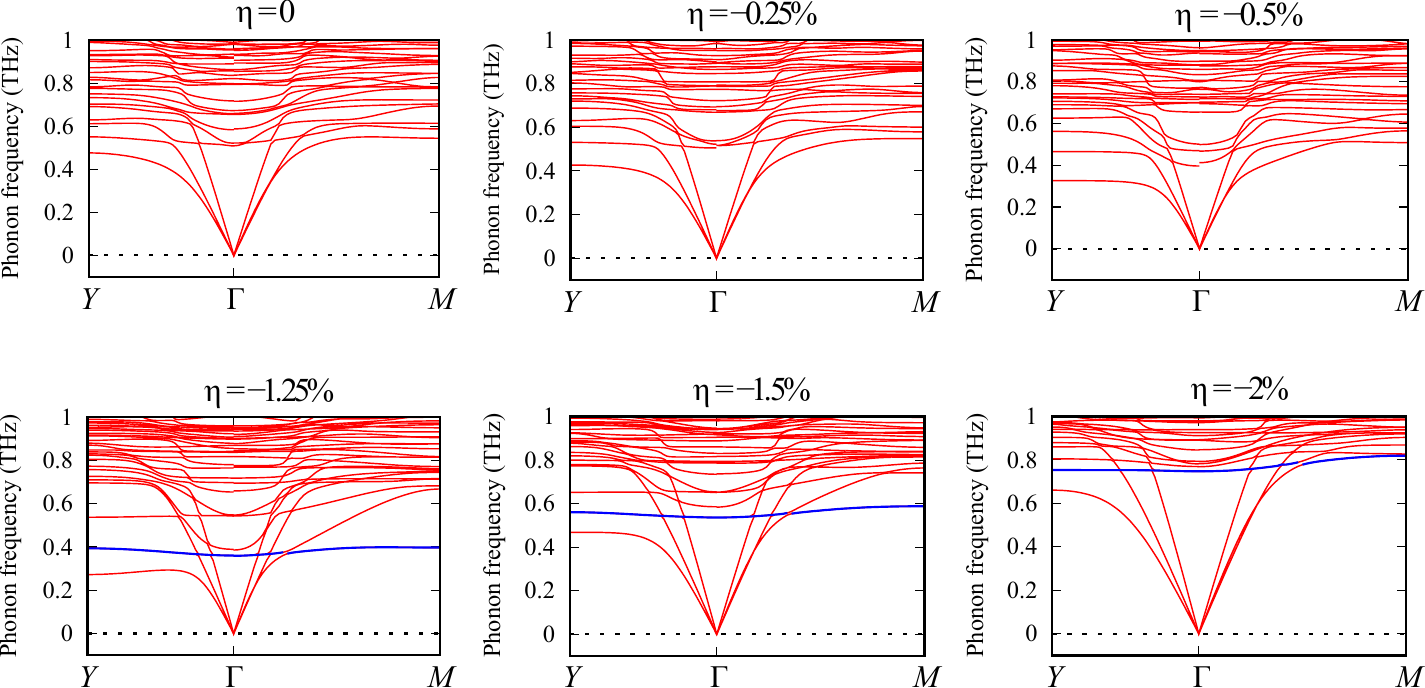}
    \caption{Evolution of the phonon band structure with epitaxial strain $\eta$ in the 6PTO/3STO superlattice at a vertical field of $\mathcal{E}_z = 1.5$ MV/cm. The chiral phonon branch is highlighted in blue.}
\label{phonon_strain_6_3}
\end{figure}

\begin{figure}[ht!]
\includegraphics[scale=1.2]{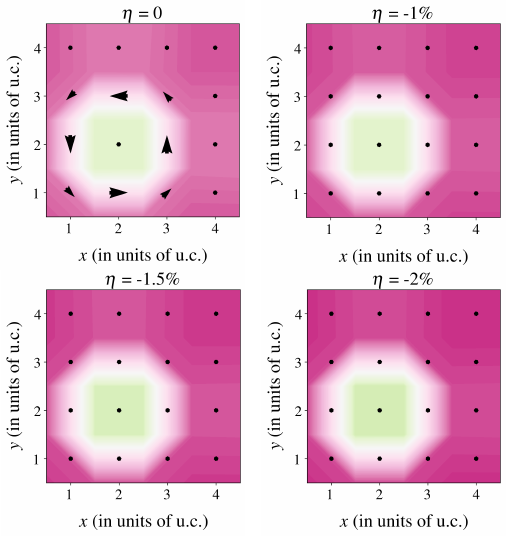}
    \caption{Evolution of the dipole field in the bulk-like PTO layer with epitaxial strain $\eta$ in the 12PTO/3STO superlattice at a vertical field of $\mathcal{E}_z = 1.5$ MV/cm.}
\label{dipole_tex_strain_12_3}
\end{figure}

\begin{figure}[ht!]
\includegraphics[scale=0.8]{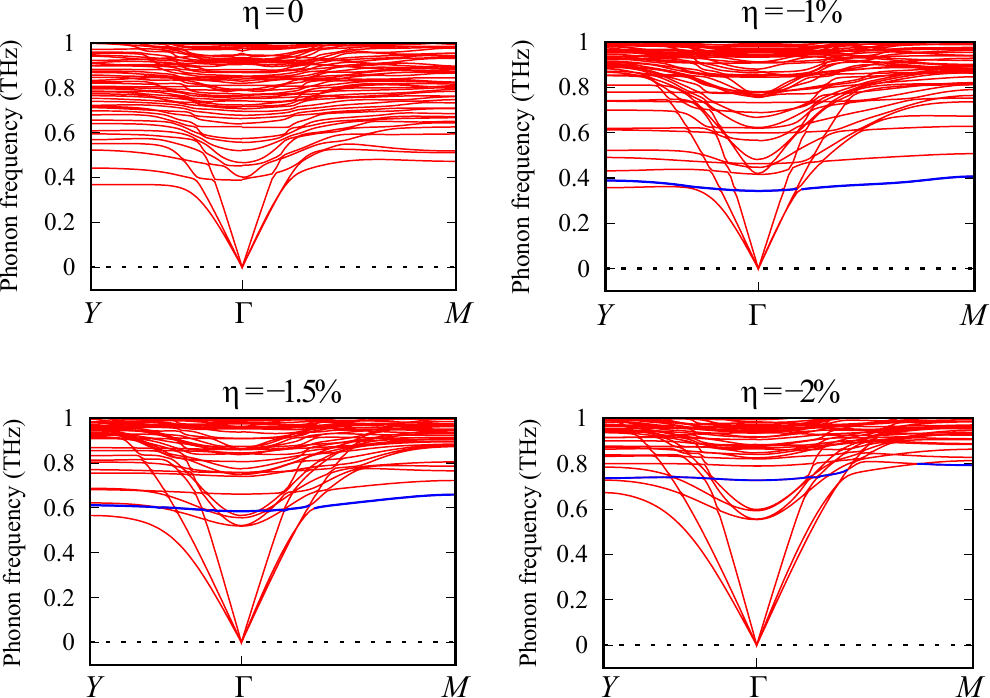}
    \caption{Evolution of the phonon band structure with epitaxial strain $\eta$ in the 12PTO/3STO superlattice at a vertical field of $\mathcal{E}_z = 1.5$ MV/cm. The chiral phonon branch is highlighted in blue.}
\label{phonon_strain_12_3}
\end{figure}

\begin{figure}[ht!]
\includegraphics[scale=0.75]{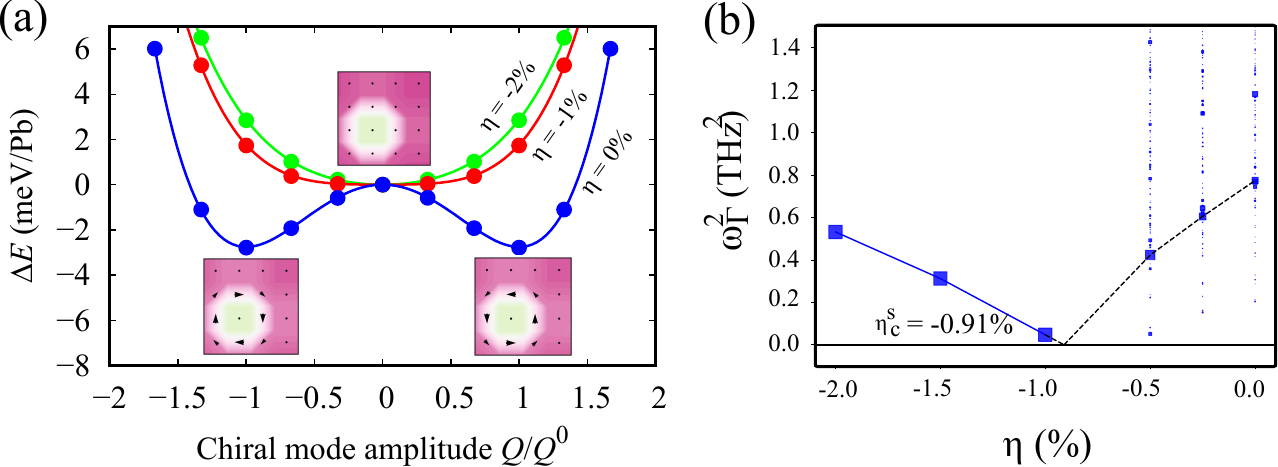}
    \caption{(a) Energy landscape of the ferroelectric bubble lattice of small bubble domains ($1 \times 1$ perovskite u.c. square section) as a function of the chiral mode amplitude $Q$ under different epitaxial strains, showing the transition from a single-well (achiral) to a double-well (chiral) potential in the tensile regime. The chiral mode amplitude is scaled to its relaxed magnitude $Q_0$ at 0\% strain. (b) Squared frequency of the chiral phonon at the $\Gamma$ point as a function of epitaxial strain, illustrating the linear softening towards the critical strain ($\eta_c$) characteristic of a soft-mode–driven phase transition. Vertical electric field is fixed at $\mathcal{E}_z = 1.5$ MV/cm.}
\label{DW_SB}
\end{figure}

\begin{figure}[ht!]
\includegraphics[scale=0.65]{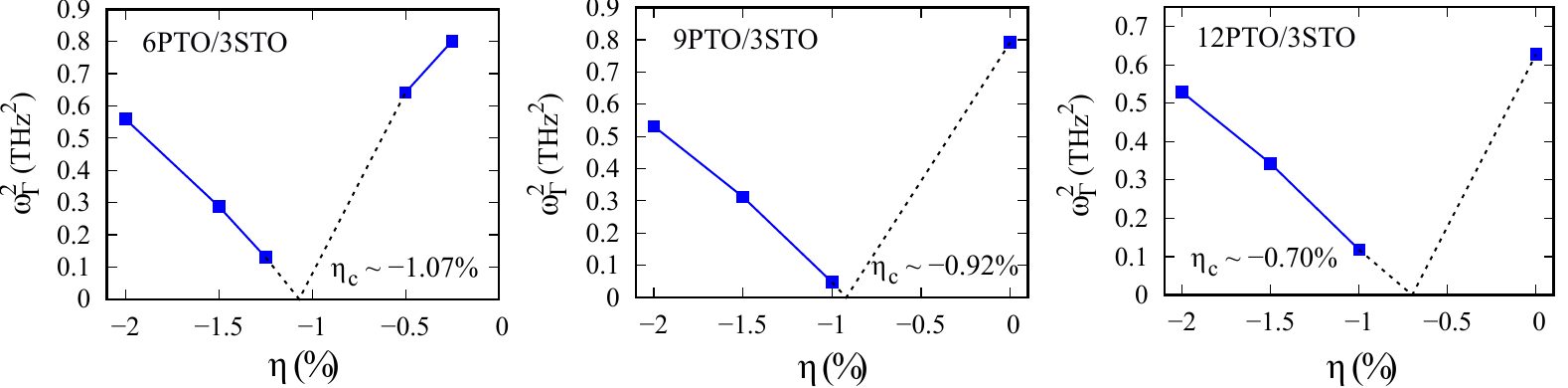}
    \caption{Squared frequency of the chiral phonon at the $\Gamma$ point as a function of epitaxial strain for different PTO layer thicknesses, showing that the strain-driven chiral phase transition is qualitatively independent of the ferroelectric layer thickness. The critical strain $\eta_c$ decreases for thicker ferroelectric layers with larger out-of-plane polarization, since a smaller compressive strain is sufficient to suppress the Bloch components of the local polarization. The vertical electric field is fixed at $\mathcal{E}_z = 1.5$ MV/cm for all calculations.}
\label{chiral_PT}
\end{figure}

\begin{figure}[ht!]
\includegraphics[scale=0.5]{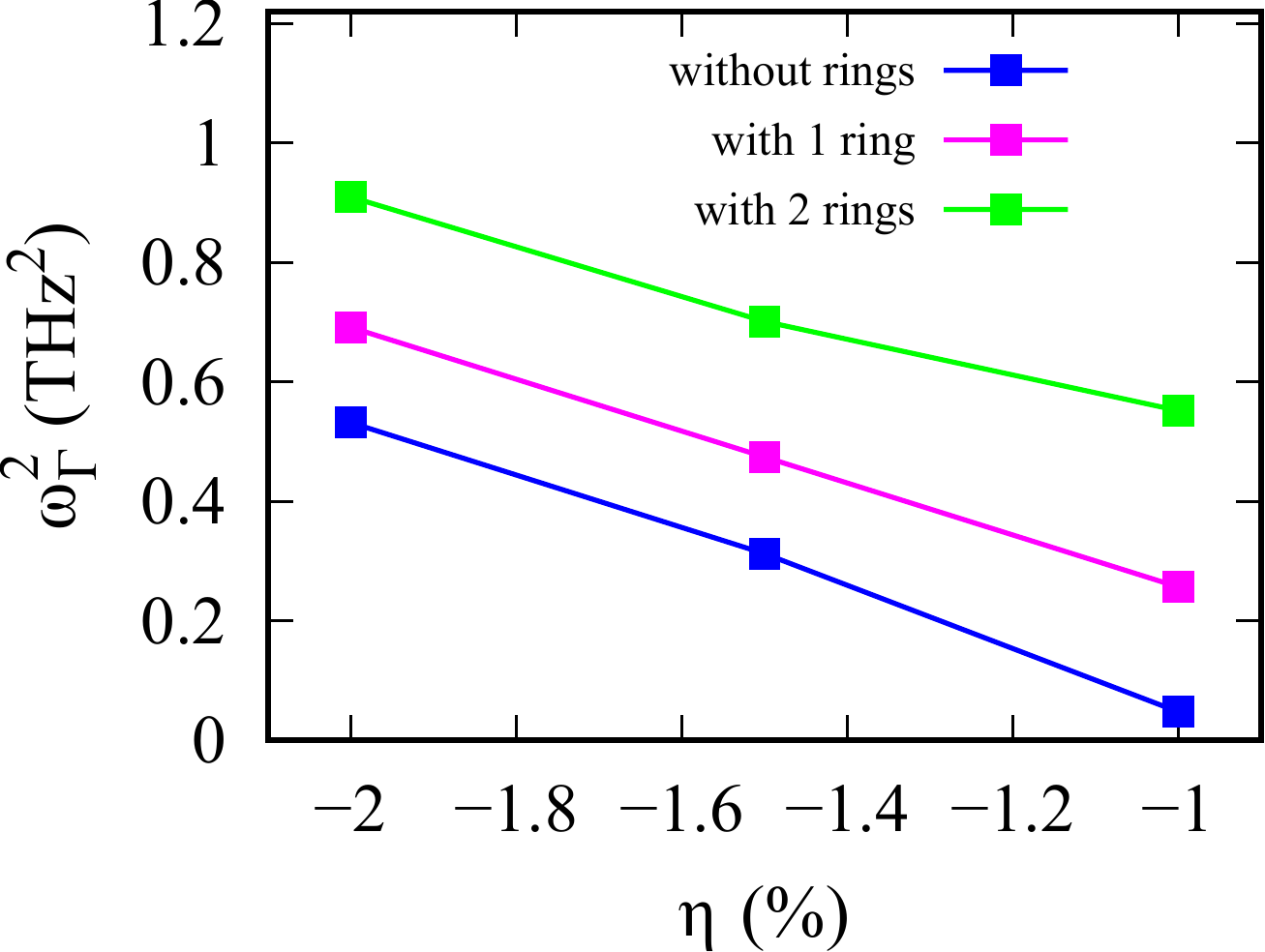}
    \caption{Evolution of the squared $\Gamma$-phonon frequencies of chiral bubbles with and without Ising rings within the PTO layers, as a function of epitaxial strain. Results are shown for the 9PTO/3STO superlattice with small bubble domains ($1 \times 1$ perovskite u.c. square section) under a vertical electric field of $\mathcal{E}_z = 1.5$ MV/cm.}
\label{nodes_field}
\end{figure}

\begin{table}[ht!]
  \setlength{\tabcolsep}{6.0pt}
  \caption{Evolution of the relative energies of ferrochiral and antiferrochiral bubble structures as a function of epitaxial strain in the 9PTO/3STO superlattice with small bubble domains ($1 \times 1$ perovskite u.c. square cross section) under a vertical electric field of $\mathcal{E}_z = 1.5$ MV/cm.}
  \label{tab-easydir}
  \centering
  \begin{tabular}{|c| c | c |c |}
    \hline
    \hline\rule{0pt}{1.0\normalbaselineskip}
   Strain $\eta$ (\%)&\multicolumn{3}{c|}{Relative energy (meV/f.u.)}\\\cline{2-4}  \rule{0pt}{1.2\normalbaselineskip}
 &Ferrochiral& Antiferrochiral along $y$&Antiferrochiral along $xy$\\
     \hline \rule{0pt}{1.2\normalbaselineskip}   
    $0.0$&0.000 &-0.099&-0.344\\\rule{0pt}{1.2\normalbaselineskip}   
    $-0.25$&0.000&-0.013&-0.141\\\rule{0pt}{1.2\normalbaselineskip}   
    $-1.0$&0.000 &0.000&0.000\\ 
    \hline 
    \hline
\end{tabular}
\end{table}

\clearpage
\newpage

\subsection*{S2. Effect of External Electric Field}
\begin{figure}[ht!]
\includegraphics[scale=0.6]{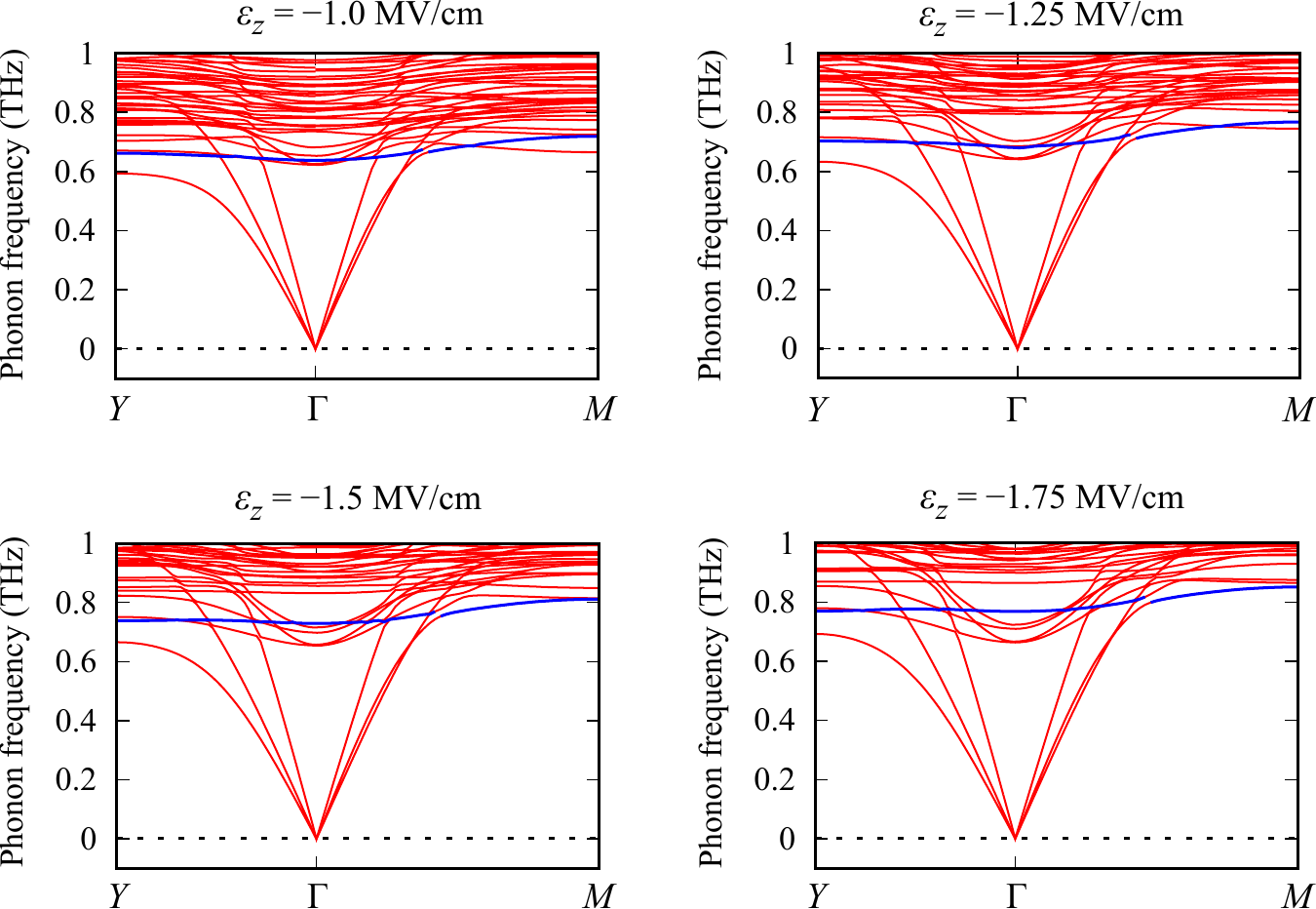}
    \caption{Evolution of the phonon spectrum with the vertical electric field in the 9PTO/3STO superlattice with small bubble domains ($1 \times 1$ perovskite u.c. square section) at $\eta  = -2\%$ strain. The chiral phonon branch is highlighted in blue.}
\label{phonon_field}
\end{figure}

\begin{figure}[ht!]
\includegraphics[scale=0.4]{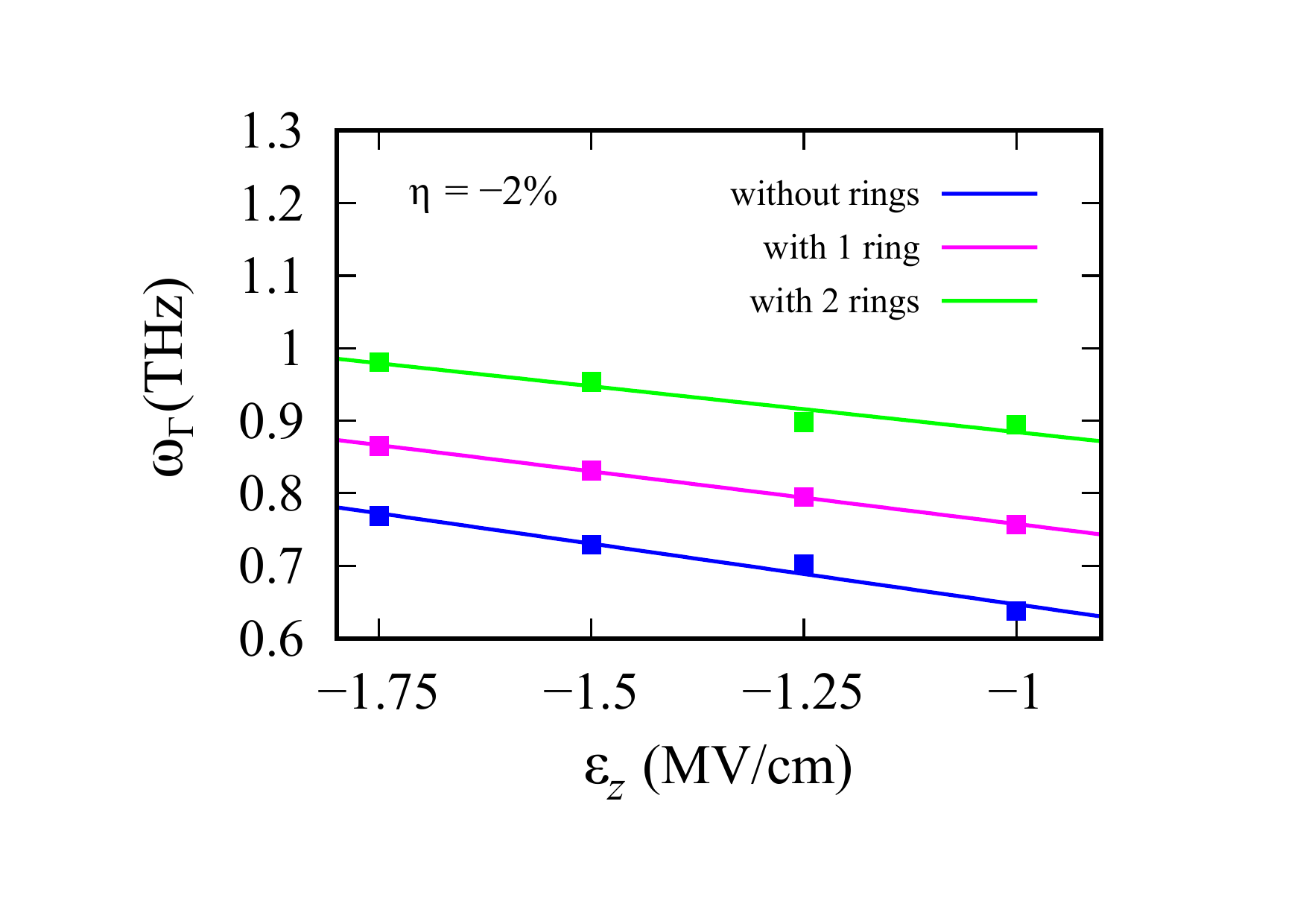}
    \caption{$\Gamma$-point phonon frequencies of chiral bubbles with and without Ising rings along the PTO layers, as a function of the vertical electric field. Results are shown for the 9PTO/3STO superlattice with small bubble domains ($1 \times 1$ perovskite u.c. square section) at $\eta = -2\%$ strain.}
\label{nodes_field_2}
\end{figure}

The linear dependence of the chiral phonon frequencies on the vertical electric field can be understood from a simple Landau free-energy model. The chirality of the electric bubbles described in this work arises from the in-plane Bloch components. To capture the effect of the external field, we therefore consider the free-energy terms that depend on the in-plane polarizations $P_x$ and $P_y$:

\begin{equation*}
\mathcal{F} \propto a (P_x^2 + P_y^2) + b^\prime (P_x^2 + P_y^2) P_z^2
\end{equation*}

which effectively includes, to lowest order, the coupling between the in-plane and out-of-plane polarization components. Then, considering the linear dependence of $P_z$ on $\mathcal{E}_z$, we obtain:
\begin{equation*}
\begin{split}
\mathcal{F}&\propto a (P_x^2 + P_y^2) + b (P_x^2 + P_y^2)\mathcal{E}_z^2 \\
&= \bigl(a + b\mathcal{E}_z^2\bigr)(P_x^2 + P_y^2), \quad a < 0 ; b > 0,
\end{split}
\end{equation*}

which shows that the energy cost of developing an in-plane polarization  varies linearly with the square of the electric field along $z$. Because this energy term defines the stiffness of the in-plane polarization, the coefficient of the quadratic term corresponds to the effective force constant, which depends on the square of the phonon frequency. This explains the linear dependence of the chiral phonon frequencies on the vertical electric field in Fig.~\ref{nodes_field_2}. 

\clearpage
\newpage
\subsection*{S3. Type-II Bubbles}\label{type2}
\begin{figure}[ht!]
\includegraphics[scale=1.5]{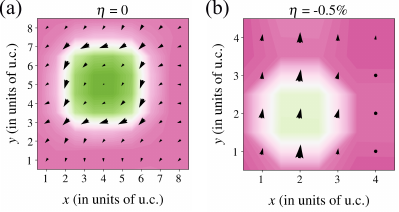}
    \caption{Dipole field in the middle PTO layer of the 9PTO/3STO superlattice. (a) Type-II bubble with a large diameter, showing a net in-plane dipole around the domain wall. (b) Type-II bubble with a smaller diameter, stabilized at $\eta = -0.5\%$ under an external field of $\mathcal{E}_z = 1.5$~MV/cm.}
\label{type_2_MC}
\end{figure}

By analyzing other phonon excitations of the achiral bubble, we identify an optical mode close in frequency to the chiral phonon at the $\Gamma$ point as a type-II bubble, i.e., an excitation with a net in-plane dipole around the domain wall, as shown in Fig.~\ref{type_2_MC}. For the large bubble with a $3 \times 3$ u.c. square section, the type-II structure is found to be around $0.7$ meV/Pb higher in energy than the corresponding uniform chiral bubble at $\eta = 0\%$. On the other hand, the type-II bubble with a $1 \times 1$ u.c. square section cannot be stabilized $\eta = 0\%$. At $\eta = -0.5\%$, it is found to be slightly more stable (by $\sim 0.2$ meV/Pb) than the corresponding small chiral bubble under an external field of $\mathcal{E}_z = 1.5$~MV/cm.

\begin{figure}[ht!]
\includegraphics[scale=0.7]{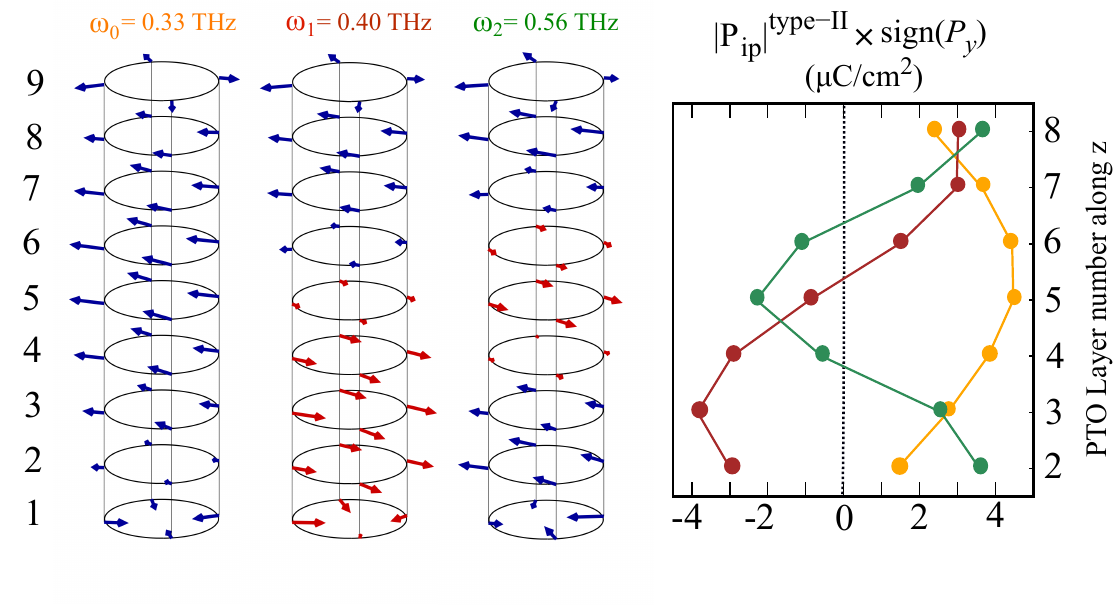}
    \caption{Vertically modulated type-II bubble structures for bubbles of small diameter. (a) Type-II bubble with uniform in-plane polarization. (b–c) Vertically modulated type-II bubbles with one and two Ising rings within the PTO layers, respectively. Dipole field only at the domain walls is shown here. (d) Layer-resolved magnitude of the in-plane polarization, analogous to the harmonic modes of a string fixed at both ends. The layers at the interface with STO [1 and 9 in (a)] layers are excluded from the figure because of their large N{\'e}el-like components.}
\label{type_2_fig}
\end{figure}

Interestingly, we also identify higher-frequency phonon modes corresponding to type-II bubble structures, involving vertical modulations of the net in-plane polarization across the PTO layers with odd or even numbers of Ising rings (see Fig.~\ref{type_2_fig}), similar to those observed in chiral bubbles. However, these configurations appear only as phonon excitations and could not be stabilized as distinct structural phases.

\end{document}